%% file: projReport.tex
\documentclass{llncs} 
\usepackage[english]{babel} 
\usepackage{enumerate}
\usepackage{macrosDL}
\usepackage{macrosDLext}

\title{Data complexity of  answering conjunctive \\queries over $\dlf{SHIQ}$ knowledge bases}

\author{Technical Report\\Faculty of Computer Science\\Free University of Bolzano}
\institute{M. Magdalena Ortiz de la Fuente, Diego Calvanese,\\Thomas Eiter and Enrico Franconi}

\begin{document}

\maketitle

\input{overview}
\input{shiqPrels}
\input{queryPrels}

\input{complFor}
\input{algorithm}
\input{tableau}

\input{conjQuer}

\input{complexity}

\bibliographystyle{plain}
\bibliography{refs} 

\end{document}

%% file: overview.tex
\begin{abstract}
In~\cite{levy00} the authors give an algorithm for answering conjunctive queries over $\dlf{ALCNR}$ knowledge bases which is $\conp$ in data complexity. 
Their technique is based on the tableau technique for checking satisfiability in $\dlf{ALCNR}$ presented in~\cite{buchheit93decidable}. In their algorithm, the blocking conditions 
of~\cite{buchheit93decidable} are weakened in such a way that the set of models their algorithm yields suffices to check query entailment. 
The algorithm we propose consists on applying a similar technique to the tableaux algorithm in~\cite{indSHIQ}, which decides the satisfiability of $\dlf{SHIQ}$ knowledge bases. 
As a result we have an algorithm for answering conjunctive queries over $\dlf{SHIQ}$ knowledge bases that is also $\conp$ in terms of data complexity.
\end{abstract}

\section{Introduction}

The idea of using description logic (DL) knowledge bases to represent the conceptual view of data repositories is becoming popular nowadays. In the context of large data repositories with a fixed schema, query answering becomes a key issue and the size of the data is the main parameter for measuring complexity. 
While atomic queries (A-Box reasoning) have always been considered an essential reasoning task in description logics, conjunctive queries and other kind of queries have recently become a topic of interest.
Data complexity of query answering over DL knowledge bases was already studied in~\cite{Scha94b}.
Many of the existing results correspond to the fragment of DLs for which the problem remains polynomial and the $\logspace$ boundary of such logics, that has been studied in detail in~\cite{calvanese}.
It is known that for rather simple DLs, even less expressive than $\dlf{ALE}$, the problem is already $\conp$ hard~\cite{Scha94b,calvanese}. 
However, results concerning complexity upper bounds are scarce. In~\cite{levy00} the authors 
prove that answering conjunctive queries over $\dlf{ALCNR}$ knowledge bases is in $\conp$ \wrt data complexity and they provide 
a worst case optimal algorithm for solving the problem.
In this work, we address the same problem for more expressive DLs, namely ones that have inverse roles and role hierarchies. 
In \cite{motik}, a data complexity $\conp$ upper bound for ground atomic queries over $\dlf{SHIQ}$ knowledge bases is given, but their technique does not yield such an upper bound for conjunctive queries. 

In this work we use a tableau algorithm. 
The algorithm proposed in~\cite{levy00} is based on the tableau technique for checking satisfiability in $\dlf{ALCNR}$ presented in~\cite{buchheit93decidable}. The key issue is that the blocking conditions of~\cite{buchheit93decidable} are weakened in such a way that it can be ensured that the query is entailed by the knowledge base iff it is entailed by the models obtained via this algorithm.
The algorithm we propose consists basically on applying the same technique to the tableaux algorithm in~\cite{indSHIQ}, which decides the satisfiability of a $\dlf{SHIQ}$ knowledge base. 
As a result we have an algorithm for answering conjunctive queries over $\dlf{SHIQ}$ knowledge bases that is CoNP in data complexity.

%% file: shiqPrels.tex
\section{Preliminaries}

\subsection{$\dlf{SHIQ}$ Knowledge Bases}

The syntax and semantics of $\dlf{SHIQ}$ are defined in the standard way. 

\begin{definition}[$\dlf{SHIQ}$ knowledge base]
Let $\mathbf{C}$ be a set of concept names and 
$\mathbf{R}$ a set of role names with a subset  
$\mathbf{R}_{+} \subseteq \mathbf{R}$ of transitive role names. 
The set of roles is $\mathbf{R} \cup \set{R^{-} \st R \in \mathbf{R}}$. 
The function $\mathsf{Inv}$ and $\mathsf{Trans}$  are defined on roles. 
$\mathsf{Inv}$ is defined as $\inverse{R} = R^{-}$ and $\inverse{R^{-}} = R$ for any role name $R$.
$\mathsf{Trans}$ is a boolean function, $\ssfunc{Trans}{R} = true$ iff $R \in \mathbf{R}_{+}$ or $\inverse{R} \in \mathbf{R}_{+}$. 

A \emph{role inclusion} axiom is an expression of the form $R \SUBS S$ where $R$ and $S$ are roles.
A \emph{role hierarchy} is a set of role inclusion axioms. 
The relation $\SUBSTRANS$ denotes the transitive closure of $\SUBS$ over a role hierarchy $\krbox{R} \cup \set{\inverse{R} \SUBS \inverse{S} \st R \SUBS S \in \krbox{R}}$.  
We say that $R$ is a \emph{sub-role} of $S$ when $R \SUBSTRANS S$, 
and a  \emph{super-role} of $S$ when $S \SUBSTRANS R$.
We will assume that it is never the case that $R$ is  both a sub-role and a super-role of $S$\footnote{This consideration is done for practical purposes, however it does not restrict the expressiveness of the language. It is clear that if $R$ is at the same time a sub-role and a super-role of $S$ both roles will have the same extension and one of them can be eliminated by replacing it by the other.}.
A role is \emph{simple} if its neither transitive nor has transitive sub-roles. 

The set of $\dlf{SHIQ}$ concepts is the smallest set such that:
\begin{itemize}
   \item
	Every concept name is a concept,
   \item
	If $C$ and $D$ are concepts, $R$ is a role, $S$ is a simple role and $n$ is a non-negative integer, then
	$C \AND D$, $C \OR D$, $\NOT C$, 
	$\ALL{R}{C}$, $\SOME{R}{C}$, $\ATLEAST{n}{S}{C}$, $\ATMOST{n}{S}{C}$
	are concepts.
\end{itemize} 

A \emph{concept inclusion axiom} is an expression of the form $C \SUBS D$ for two concepts $C$ and $D$. 
A \emph{terminology} or T-Box is a set of concept inclusion axioms.

Let $\mathbf{I}$ be a set of individual names. 
An \emph{assertion} is an expression that can have the form 
$C(a)$, $R(a,b)$ or $a \NEQ b$ 
where $C$ is a concept, $R$ is a role and $a,b \in \mathbf{I}$.
An \emph{A-Box} is a set of assertions.

A  $\dlf{SHIQ}$ knowledge base is a triple $K = \tup{\krbox{A}, \krbox{R}, \krbox{T}}$, where  
$\krbox{A}$ is an A-Box,  
$\krbox{R}$ is role hierarchy and
$\krbox{T}$ is a terminology. 
\end{definition}

The semantics of $\dlf{SHIQ}$ knowledge bases is given by interpretations. 

\begin{definition}[Interpretation]\label{defInterpret}
An interpretation $\I = \inter{\I}$ is defined for 
a set of individual names $\mathbf{I}$,
a set of concepts $\mathbf{C}$ and
a set of roles $\mathbf{R}$.
The set $\dom{\I}$ is called \emph{domain} of $\I$.
The valuation $\Int{\I}{\cdot}$ maps 
each individual name in $\mathbf{I}$ to an element in $\dom{\I}$,
each concept in $\mathbf{C}$  to a subset of $\dom{\I}$,
and each role in $\mathbf{R}$ to a subset of $\dom{\I} \times \dom{\I}$.
Additionally, for any concepts $C$, $D$,
any role $R$ and any non-negative integer $n$, 
the valuation $\Int{\I}{\cdot}$ must satisfy the following equations:
\begin{displaymath}
    \begin{array}{rcl}
	\Int{\I}{R}		& = &	(\Int{\I}{R})^{+}  \text{ \ \ for each role } R \in \mathbf{R}_{+}\\
	\Int{\I}{(R^{-})}	& = &	\set{\tup{y,x} \st \tup{x,y} \in \Int{\I}{R}}		\\
	\Int{\I}{(C \AND D)}	& = &	\Int{\I}{C} \cap \Int{\I}{C}				\\
	\Int{\I}{(C \OR D)}	& = &	\Int{\I}{C} \cup \Int{\I}{C}				\\
	\Int{\I}{(\NOT C)}	& = &	\dom{\I} \setminus \Int{\I}{C}				\\
	\Int{\I}{(\ALL{R}{C})}	& = &	\set{x \st \text{ for all $y$, }  \tup{x,y} \in \Int{\I}{R} \text{ implies } y \in \Int{\I}{C}}\\
	\Int{\I}{(\SOME{R}{C})}	& = &	\set{x \st \text{ for some $y$, } \tup{x,y} \in \Int{\I}{R} \text{ and } y \in \Int{\I}{C}}    \\
	\Int{\I}{(\ATLEAST{n}{R}{C})}	& = &	\set{x \st \card{\set{y \st \tup{x,y} \in \Int{\I}{R} \text{ and } y \in \Int{\I}{C}}} \geq n }  \\
	\Int{\I}{(\ATMOST{n}{R}{C})}	& = &	\set{x \st \card{\set{y \st \tup{x,y} \in \Int{\I}{R} \text{ and } y \in \Int{\I}{C}}} \leq n }  
    \end{array}
\end{displaymath}
\end{definition}

\begin{definition}[Model of a knowledge base]
An interpretation $\I$ satisfies an assertion $A$ iff:
\begin{displaymath}
    \begin{array}{rl}
	a \in \Int{\I}{C} 		& \text{ \ \ if $A$ is of the form } C(a) \\
	\tup{a,b} \in \Int{\I}{R} 	& \text{ \ \ if $A$ is of the form } R(a,b) \\
	\Int{\I}{a} \neq \Int{\I}{b} 	& \text{ \ \ if $A$ is of the form } a \NEQ b \\
    \end{array}
\end{displaymath}
An interpretation $\I$ satisfies an A-Box $\krbox{A}$ if it satisfies every assertion in $\krbox{A}$.
$\I$ satisfies a role hierarchy $\krbox{R}$ if $\Int{\I}{R} \subseteq \Int{\I}{S}$ for every $R \SUBS S$ in $\krbox{R}$.
$\I$ satisfies a terminology $\krbox{T}$ if $\Int{\I}{C} \subseteq \Int{\I}{D}$ for every $C \SUBS D$ in $\krbox{T}$.
$\I$ is a model of $K = \tup{\krbox{A}, \krbox{R}, \krbox{T}}$ if it  satisfies $\krbox{A}$, $\krbox{R}$ and $\krbox{T}$.
\end{definition}

\input{axiomSys}

%% file: axiomSys.tex
A $\dlf{SHIQ}$ concept is said to be in \emph{negation normal form} (NNF)  if negation occurs only in front of concept names. Since concepts can be translated into NNF in linear time~\cite{indSHIQ}, we will assume that all concepts are in NNF. We denote by $NNF(\NOT C)$ the NNF of the concept $\NOT C$.
The closure of a concept $\clos{C}$ is the smallest set  containing $C$ that is closed under subconcepts and negation (in NNF).
For a knowledge base $K$, $\clos{K} = \cup_{C(a) \in K} \clos{C}$.

\subsubsection{Global Constraint Concepts}.

A knowledge base $K$ has an associated set of concepts that we will call the \emph{global constraint concepts} of $K$. This set contains two kinds of concepts:
\begin{itemize}
\item
For each concept inclusion axiom $C \SUBS D$ in the TBox, there is a global constraint concept of the form $\NOT C \OR D$. This way, if we assure that all individuals in a model belong to the extension of global constraint concepts, the model will satisfy the T-Box of $K$\footnote{In~\cite{indSHIQ} the authors consider an internalised T-Box. We do not make this assumption.}. 

\item
We will consider that $K$, additionally to the A-Box, T-Box, R-Box, might have a set of \emph{distinguished concepts} names that we will denote $\C_K$. In order not to make the notation too cumbersome, we will not denote it explicitly as a part of $K$.
For all concept names $C$ in $\C_K$ the concept $C \OR \NOT C$ belongs to the global constraints of $K$. 
In the algorithm we present in the following sections, we will use partial representations of models of a knowledge base to verify whether some formula $Q$ is entailed in them. 
In these partial representations it may remain undecided whether some individuals belong to the extension of a concept or of its negation. However, for the concepts that appear in $Q$, we want to assure that the decision is taken. We will later see that in our framework, the set $C_K$ will be used to represent the concepts that may appear in the queries to be answered \footnote{If $C_K = \clos{K}$, the algorithm can be used to check entailment w.r.t. any concept in the knowledge base, however this may be inconvenient from an implementation perspective.}. 
\end{itemize}

\begin{definition}[Global Constraint Concepts]
Given a knowledge base $K = \kb{A}{T}{R}$ and a set of distinguished concept names $\C_K$, 
the set of \emph{global constraint concepts for $K$ and $\C_K$} is defined as
$\axioms{K,\C_K} = 	\set{\NOT C \OR D \st C \SUBS D \in \krbox{T}} 
		\cup 	\set{C \OR \NOT C \st C \in \C_K}$.
\end{definition}

If not stated otherwise, in the following 
$K$ will denote a $\dlf{SHIQ}$ knowledge base $K = \tup{\krbox{A},\krbox{R},\krbox{T}}$,
$\roles{K}$ the roles occurring in $K$ together with their inverses,
$\clos{K}$ the closure of the concept names occurring in $\krbox{A}$,
$\C_K$ will denote a distinguished set of concept names,
and $\indivs{K}$ the individual names occurring in $\krbox{A}$.

%% file: queryPrels.tex
\subsection{Answering Conjunctive Queries over Knowledge Bases}

In the traditional database setting, free variables in a query are called distinguished variables. For a query $Q$ that has $\vect{X}$ as distinguished variables, the query answering problem over $K$ consists on finding all the possible tuples of constants $\vect{T}$ of the same arity as $\vect{X}$ such that when $\vect{X}$ is substituted by $\vect{T}$ in $Q$, it holds that $K \models Q$. The set of such tuples $\vect{T}$ is the answer of the query.
Query answering has an associated recognition problem: given a tuple $\vect{T}$, the problem is to verify whether $\vect{T}$ belongs to the answer of $Q$\footnote{This problem is usually known as the \emph{query output problem}.}.
We say that query answering for a certain description logic is in a class $C$ \wrt data complexity when the corresponding recognition problem is in $C$. Since we will only focus on the recognition problem, we allow conjunctive queries to contain constants and we are assuming that all variables in the query are existentially quantified. 

\begin{definition}[conjunctive query]
A \emph{conjunctive query} over a knowledge base $K$ is a sentence of the form 
	\begin{displaymath}
		(\exists \vect{Y}).p_1(\vect{Y_1})\ld{\land}p_n(\vect{Y_n})
	\end{displaymath}
where $p_1 \cld p_n$ are either roles in $\roles{K}$ or concepts in $\C_{K}$;
$\vect{Y_1} \cld \vect{Y_n}$ are tuples of variables and constants.
$V_Q = \vect{Y_1} \ld{\cup} \vect{Y_n}$ denotes the set of variables and constants in $Q$.
The set of literals in $Q$ is $L_Q = \set{p_1(\ol{Y_1}) \cld p_n(\ol{Y_n})}$, and the cardinality of $L_Q$ will be denoted by $n_Q$.
\end{definition}

Conjunctive queries are interpreted in the standard way, i.e. $\I = \inter{\I}$ is a model of $Q$ if there is a mapping $\sigma$ from the variables and constants in $Q$ to objects in $\dom{\I}$ such that $\sigma$ is the identity on all constants and $\sigma(\vect{Y}) \in \Int{\I}{p}$ for all $p(\vect{Y}) \in L_Q$.
For a knowledge base $K$ and a query $Q$, we say that $K \models Q$ iff for every interpretation $\I$, $\I \models K$ implies $\I \models Q$.
Analogously, for a completion forest $\F$ and a query $Q$, we say that $\F \models Q$ iff for every interpretation $\I$, $\I \models \F$ implies $\I \models Q$.

\begin{definition}[Conjunctive Query Entailment]
Let $K$ be a knowledge base and let $Q$ be conjunctive query. 
The \emph{conjunctive query entailment} problem is to decide whether $K \models Q$. 
\end{definition}

We are interested in solving the conjunctive query entailment problem. 
It is important to notice that the conjunctive query entailment problem is not reducible to satisfiability of knowledge bases, since the negation of the query can not be expressed as a part of a knowledge base. 
For this reason, the known algorithms for reasoning over knowledge bases do not suffice. 
A knowledge base $K$ has an infinite number possibly infinite models, and we have to verify whether the query $Q$ is entailed by all of them. 
In general, we want to provide an entailment algorithm, i.e. an algorithm for checking whether a sentence $Q$ with a particular syntax (namely, a conjunctive query) is entailed by a $\dlf{SHIQ}$ knowledge base $K$.
Informally, our algorithm differs from the one proposed in~\cite{indSHIQ} for reasoning with individuals in $\dlf{SHIQ}$ in the fact that, since they only focus on problems that can be reduced to checking satisfiability, they only need to ensure that if the knowledge base has some model then their algorithm will obtain a model.
In our case, however, this is not enough. We need to make sure that the algorithm obtains a set of models $M$ such that $Q$ is entailed by $K$ iff it is entailed by every model in $M$.

%% file: complFor.tex
\section{A $\dlf{SHIQ}$ Entailment Algorithm}

We will provide an algorithm for checking entailment of some sentence $Q$ in a $\dlf{SHIQ}$ knowledge base $K$, i.e. to check if all models of $K$ are models of $Q$.  
Like the algorithm in~\cite{indSHIQ}, we will use \emph{completion forests}. 
A completion forest is a relational structure that captures sets of models of a knowledge base. 
A completion forest is always finite, and it represents a set of possibly infinite models.
When defining completion forests, we will use a parameter $n$ that is not present in~\cite{indSHIQ}. This parameter will be 
crucial in ensuring that the application of our algorithm will yield a set of models $M$ such that $Q$ is entailed by $K$ iff it is entailed by every model in $M$. We will see later that this parameter will take values that depend on $Q$.

\subsection{Completion Forests}

A forest will be defined as a set of variable trees. 
A variable tree is a tree where the nodes are variables, and where the nodes and arcs of the tree are labeled. For any nodes $n_1$ and $n_2$, $\labelfunc{n_1}$ will denote the label of $n_1$ and 
$\labelfunc{\tup{n_1,n_2}}$ will denote the label of the arc that goes from $n_1$ to $n_2$. 

\begin{definition}[$n$-tree equivalence]
Given a variable tree $V$ s.t. $v$ is a node of $V$, the \emph{$n$-tree of $v$} is the subtree of $V$ that has $v$ as its root and contains the successors of $v$ that are at most $n$ direct successor arcs away.
We denote by $V_n(v)$ the set of nodes of $V$ that appear in the $n$-tree of $v$. 
Two variables $v$, $w$ in $V$ are \emph{$n$-tree equivalent in $V$} if there is an isomorphism $\psi$ between their $n$-trees, i.e. $\psi : V_n(v) \ra V_n(w)$ is a mapping such that:
\begin{itemize}
\item
$\psi(v) = w$
\item
for every node $n$ in $V_n(v)$, $\labelfunc{n} = \labelfunc{\psi(n)}$ 
\item
for every arc connecting two nodes $n_1$ and $n_2$ in $V_n(v)$, \\
$\labelfunc{\tup{n_1,n_2}} = \labelfunc{\tup{\psi(n_1),\psi(n_2)}}$. 
\end{itemize}

\end{definition}

\begin{definition}[$n$-Witness]
Let $V$ be a variable tree where both $v$ and $w$ are nodes. 
We say that $w$ is an \emph{$n$-witness of $v$ in $V$} iff  $w$ is an ancestor of $v$ in $V$,  $w$ is $n$-tree equivalent to $v$ in $V$ and $v$ is not in the $n$-tree of $w$.
Let $t$ denote the $n$-tree of which $v$ is root, $t'$ the $n$-tree that has $w$ as root, and let $\psi$ denote an isomorphism between $t$ and $t'$. In this case, we say that $t'$ tree-blocks $t$. For all variables $x$ in $t$, we say that $\psi(x)$ tree-blocks $x$. 
\end{definition}

\begin{definition}[$n$-Completion Forest]

A \emph{completion forest} for a knowledge base $K$ is given by a forest of trees and an inequality relation $\NEQ$ which is assumed to be symmetric. 
The forest is a set of variable trees whose roots are the individuals in $\indivs{A}$. The roots can be connected by edges in an arbitrary way.
$\labelfunc{x} \subseteq \clos{K}$  denotes the label of a node $x$, 
and $\labelfunc{\tup{x,y}} \subseteq \roles{K}$  denotes the label of an edge $\tup{x,y}$.
 
If two nodes $x$, $y$ are connected by an edge with $R \in \tab{L}(\tup{x,y})$ and $R \SUBSTRANS S$ then $y$ is an \emph{$S$-successor} of $x$ and $y$ is an \emph{$\inverse{S}$-predecessor} of $x$. 
If $y$ is an $S$-successor of $x$, then $y$ is an  \emph{$S$-descendant} of $x$.
If $z$ is an $S$-descendant of $x$, $y$ is an $S$-descendant of $z$ and $S \in \mathbf{R}_+$, then $y$ is an  \emph{$S$-descendant} of $x$.
If $x$ is an $S$-successor or an $\inverse{S}$-predecessor of $y$, then $x$ is an \emph{$S$-neighbor} of $y$.
If $x$ is an $S$-successor of $y$ for some role $S$, then $x$ is a \emph{successor} of $y$ and $y$ is a \emph{predecessor} of $x$.  
The transitive closure of predecessor is called \emph{ancestor}.

A node is \emph{blocked} iff it is not a root node and it is either directly or indirectly blocked.
A node is \emph{indirectly blocked} iff one of its ancestors is blocked or if it's a successor of a node $x$ and $\tab{L}{\tup{x,y}} = \emptyset$.
A node is \emph{directly blocked} iff none of its ancestors are blocked and it is a leaf of an $n$-tree that is tree-blocked. 
\end{definition}

\begin{definition}[Clash free completion forest]
A node $x$ in a completion forest $\F$ contains a \emph{clash} iff 
 for some concept name $C$, $C \in \tab{L}(x)$ and $\NOT C \in \tab{L}(x)$ or 
if $\ATMOST{n}{R.C}  \in \tab{L}(x)$ and $x$ has $n+1$ R-successors $y_0 \cld y_n$ such that $C \in  \tab{L}(y_i)$ for all $y_i$ and  $y_i \neq y_j \in \F$ for all $0 \leq i \lneq j \leq n$.
A completion forest $\F$ is \emph{clash free} if none of its nodes contains a clash in $\F$.
\end{definition}

\begin{definition}[Complete completion forest]
A completion $\F$ is complete if none of the rules in Table~\ref{expRules} can be applied to it. 
\end{definition} 

%% file: algorithm.tex
\subsection{The Completion Forest Algorithm}

Given a knowledge base $K = \kb{A}{R}{T}$ and a blocking parameter $n$, the algorithm does the following:
An initial completion forest for $K$ is built and it is expanded using the rules in Table~\ref{expRules} until no more expansions can be obtained.
The (possibly empty) set of complete and clash-free $n$-completion forests obtained by this expansion induce a set of models for $K$.
As we will see in the coming sections, this set of models can be used to check entailment of a conjunctive query $Q$ if  a suitable $n$ (depending on $Q$) is used.

\subsubsection{Initializing the Completion Forest.}

An initial completion forest $\F_K$ for a knowledge base $K$ is constructed as follows:
\begin{itemize}
\item
For each individual $a_i \in \indivs{K}$ a node $a_i$ is introduced.
\item
An edge $\tup{a_i,a_j}$ is created iff $R(a_i,a_j) \in \krbox{A}$ for some role $R$.
\item
The labels of these nodes and edges as well as the $\NEQ$ relation are initialized as follows:
   \begin{center}
	\begin{tabular}{ll}
	    $\tab{L}(a_i)$ 		& $:= \set{C \st C(a_i) \in \krbox{A}} \cup \axioms{K,\C_K}$ \\
   	    $\tab{L}(\tup{a_i,a_j})$& $:= \set{R \st R(a_i,a_j) \in \krbox{A}}$\\
	    $ a_i \neq a_j$		& iff $a_i \neq a_j \in \krbox{A}$ 
	\end{tabular}
   \end{center}
\end{itemize}

\subsubsection{Expanding the Completion Forests.}

From the initial completion forest, new completion forests for $K$ can be obtained by applying the rules in Table~\ref{expRules}.
Note that the application of the rules is non-deterministic. Different choices for $E$ in the $\OR$-rule and the \emph{choose}-rule generate different forests. 
The $\exists$-rule and the $\geq$-rule are called \emph{generating rules} since they add new nodes to the forest.

\begin{table}
	\begin{center}
		\input{tableExpRules}
	\end{center}
	\caption{Expansion Rules}
	\label{expRules}
\end{table}

The set of $n$-completion forests for a knowledge base $K$ is denoted by $\nforests{F}{K}{n}$ and it is the smallest set satisfying the following conditions:
\begin{enumerate}
     \item
	The initial completion forest $\F_{K}$ is a completion forest for $K$.
     \item
	If $\F$ is a legal $n$-completion forest for $K$ 
	and $\F'$ can be obtained from  $\F$ by applying one of the rules in Table~\ref{expRules} using $n$-blocking, 
	then $\F'$ is a $n$-completion forest for $K$.
\end{enumerate}

\subsubsection{Completion Forests as Semantic Objects.}

Semantically, we can interpret a completion forest in the way we interpret a knowledge base.
For a knowledge base $K$ and a completion forest $\F$ for $K$, 
note that all the individuals in $\indivs{K}$ are nodes in $\F$,
node labels in $\F$ are concepts in $\clos{K} \cup \C_K$ and
edge labels in $\F$ are roles in $\roles{K}$,
hence interpretations for $K$ can be interpretations for $\F$ and vice-versa.
We will see completion forests as  a representation of a set of models of the knowledge base. 
It is not a common practice to give a semantical interpretation to completion forests. However, this reading will make easier  some of our results and proofs.

\begin{definition}[Model of a completion forest]\label{defModCompl}
For an $n$-completion forest $\F$ for $K$, 
$\F \in \nforests{F}{K}{n}$,
an interpretation $\I = \inter{\I}$ is a model of $\F$, represented $\I \models \F$
if $\I \models K$ and 
for all nodes $x, y \in \F$ the following hold:
    \begin{itemize}
	\item
	if $C \in \labelfunc{x}$, then $\Int{\I}{x} \in \Int{\I}{C}$
	\item
	if $R \in \labelfunc{\tup{x,y}}$ then $\tup{\Int{\I}{x}, \Int{\I}{y}} \in  \Int{\I}{R}$
	\item
	if $x \NEQ y \in \F$, then  $\Int{\I}{x} \neq  \Int{\I}{y}$
    \end{itemize}  
\end{definition}


We want to emphazise that in order to be a model of a completion forest for $K$, an interpretation must be a model of $K$.
The initial completion forest is just an alternative representation of the knowledge base, and it has exactly the same models. 
When we expand the forest, we will make choices and obtain new forests that capture a subset of the models of the knowledge base.
Note that if an interpretation $\I = \inter{\I}$ is a model of $\F$, then all nodes in $\F$ will be mapped to an object in $\dom{\I}$, however there might be objects in $\dom{\I}$ that are not the image of any node in $\F$.

\begin{lemma}\label{FKiffK}
An interpretation $\I$ is a model of $\F_K$
iff $\I$ is a model of $K$.
\end{lemma}

\begin{proof}
The if direction follows from Definition~\ref{defModCompl}.
To prove the other direction, it suffices to consider an arbitrary model $\I$ of $K$ and verify that for
for all nodes $x, y \in \F_K$ the following hold:
    \begin{enumerate}[{\it (i)}]
	\item \label{itconc}
	if $C \in \labelfunc{x}$, then $\Int{\I}{x} \in \Int{\I}{C}$
	\item \label{itrole}
	if $R \in \labelfunc{\tup{x,y}}$ then $\tup{\Int{\I}{x}, \Int{I}{y}} \in  \Int{\I}{R}$
	\item\label{itineq}
	if $x \NEQ y \in \F$, then  $\Int{\I}{x} \neq  \Int{\I}{y}$
    \end{enumerate}   

By definition, the nodes in $\F_{K}$ correspond exactly to the individuals in $\indivs{K}$. 
For each of these individuals $a_i$,
the label of $a_i$ in $\F_{K}$ is given as $\labelfunc{a_i} = \set{C \st C(a_i) \in \krbox{A}} \cup \axioms{K,\C_K}$.
Since $\I$ is a model of $\krbox{A}$, 
if  $C(a_i) \in \krbox{A}$ then $\Int{\I}{a_i} \in \Int{\I}{C}$.
For any concept $C \in \axioms{K,\C_K}$, 
either $C$ is of the form $\NOT D \OR E$ for some $D \SUBS E$ in $\krbox{T}$ or 
$C$ is of the form $D \OR \NOT D$ for an arbitrary concept $D$.
In the first case, $\Int{\I}{a_i} \in \Int{\I}{(\NOT D \OR E)}$ must hold because $\I$ is a model of $\krbox{T}$.
In the other case, $\Int{\I}{x} \in \Int{\I}{(D \OR \NOT D)}$ holds for any individual $x$ in $\dom{\I}$ and any concept $D$ by the definition of interpretation.
So we have that $\Int{\I}{a_i} \in \Int{\I}{C}$ for every $C \in \labelfunc{a_i}$ and item~{\it(\ref{itconc})} holds.
The label of a pair of nodes $a_i$, $a_j$ in $\F_K$ is given by $\labelfunc{\tup{a_i,a_j}} = \set{R \st R(a_i,a_j) \in \krbox{A}}$.
Since $\I$ is a model of $\krbox{A}$, 
$\tup{\Int{\I}{a_i} \Int{\I}{a_i}} \in \Int{\I}{R}$ for every $R(a_i,a_j)$ in $\krbox{A}$, hence item~{\it(\ref{itrole})} holds.
Analogously, the $\NEQ$ relation was initialized with $a_i \neq a_j$ for every $a_i \NEQ a_j$ in $\krbox{A}$, 
so  item~{\it(\ref{itineq})} will also hold for any $\I$ model of $\krbox{A}$.
\end{proof}

Finally, for a set of completion forests $\mathbb{F}$, we will denote by $\mathsf{ccf}(\mathbb{F})$ the set of forests in $\mathbb{F}$ that are complete and clash free. 
For a knowledge base $K$, the union of all the models of the forests in $\mathsf{ccf}(\nforests{F}{K}{n})$ captures all the models of $K$, as we prove in Proposition~\ref{KmodelssomeF}.
This result is crucial, since it allows us to ensure that checking the forests in $\mathsf{ccf}(\nforests{F}{K}{n})$ suffices to check all models of $K$.
In order to prove this result, we will first prove the following lemma. It states that when applying any of the rules in Table~\ref{expRules}, no models are lost. 

\begin{lemma}\label{FmodelssomeFp}
Let $\F$ be a completion forests in $\nforests{F}{K}{n}$,
let $r$ be a rule in Table~\ref{expRules}
and let $\mathbf{F}$ be the set of $n$-completion forests that can be obtained from $\F$ by applying $r$.
Then for every $\I$ such that  $\I \models \F$
there is some $\F' \in \mathbf{F}$ such that $\I \models \F'$.
\end{lemma}

\begin{proof}
We will do the proof for each rule $r$ in Table~\ref{expRules}.

First we will consider the deterministic, non-generating rules. 
There is only one $\F'$ in $\mathbf{F}$ and the models of $\F$ are exactly the models of $\F'$. 
For the case of the $\AND$-rule, there is some node $x$ in $\F$ s.t. $C_1 \AND C_2 \in \labelfunc{x}$.
Since $\I$ is a model of $\F$, then $\Int{\I}{x} \in \Int{\I}{(C_1 \AND C_2)}$, and since $\I$ is a model of $K$, then both
$\Int{\I}{x} \in \Int{\I}{C_1}$ and $\Int{\I}{x} \in \Int{\I}{C_2}$ hold. 
The inequality relation and all labels in $\F'$ are exactly as in $\F$, 
the only change is that $\set{C_1, C_2} \subset \labelfunc{x}$ in $\F'$, so $\I \models \F'$. 
 
The cases of the the $\forall$-rule and the  $\forall_{+}$-rule, are similar to the $\AND$-rule. 
All labels of $\F$ are preserved in $\F'$. 
Only the label of the node $y$ to which the rule was applied is modified, having in $\F'$ 
$C \subset \labelfunc{y}$ or 
$\ALL{S}{C} \subset \labelfunc{x}$ respectively. 
Since $\I$ is a model of $K$, 
$\Int{\I}{x} \in \Int{\I}{(\ALL{S}{C})}$ and $y$ and $S$-neighbour of $x$ imply $\Int{\I}{y} \in \Int{\I}{C}$, and 
$\Int{\I}{x} \in \Int{\I}{(\ALL{}{C})}$ and $y$ and $R$-neighbour of $x$ for some transitive sub-role of $S$ imply $\Int{\I}{y} \in \Int{\I}{(\ALL{}{C})}$, then trivially $\I \models \F'$ in both cases. 

Let us analyze the non-deterministic rules.
For the case of the $\OR$-rule,
there is some node $x$ in $\F$ s.t. $C_1 \OR C_2 \in \labelfunc{x}$.
After applying the $\OR$-rule, we will have two forests $\F_1'$, $\F_2'$
with $\set{C_1} \subset \labelfunc{x}$ in $\F_1'$ and $\set{C_2} \subset \labelfunc{x}$ in $\F_2'$ respectively. 
For every $\I$ such that $\I$ is a model of $\F$ we have $\Int{\I}{x} \in \Int{\I}{(C_1 \OR C_2)}$, and since $\I$ is a model of $K$, 
then  either $\Int{\I}{x} \in \Int{\I}{C_1}$ or $\Int{\I}{x} \in \Int{\I}{C_2}$ hold. 
If it is the case that $\Int{\I}{x} \in \Int{\I}{C_1}$, then $\I \models \F_1'$, and otherwise $\I \models \F_2'$, 
so the claim holds. 

The proof of the choose rule is trivial, since after its  application we will have two forests $\F_1'$, $\F_2'$
with $\set{C} \subset \labelfunc{x}$ in $\F_1'$ and $\set{\lxnot C} \subset \labelfunc{x}$ in $\F_2'$ respectively,
but since trivially $\Int{\I}{x} \in \Int{\I}{(C \OR \lxnot C)}$ holds for any $x$, any $C$ and any $\I$ model of $K$, 
then for every $\I$ either $\I \models \F_1'$ or $\I \models \F_2'$ holds.

When the $\leq$-rule or the $\leq_r$-rule are applied to a variable $x$ in $\F$, there are some variables $y$, $z$ neighbours of $x$
s.t. $y$ is identified with $z$ in $\F'$.
This can only be done if we do not have that $\Int{\I}{z} \neq \Int{\I}{y}$ in $\I$, 
hence it must be the case that  $\Int{\I}{z} = \Int{\I}{y}$.
In $\F'$, we will add the pair $\tup{z,y}$ to the extension of $\EQ$.
Due to $\Int{\I}{z} = \Int{\I}{y}$ the extensions of all labels of $\F$ will be preserved in $\F'$ and so $\I \models \F'$ holds.

Finally we consider the two generating rules.
For the case of the $\exists$-rule, 
since the propagation rule was applied, there is some $x$ in $\F$ such that $\SOME{R}{C} \in \labelfunc{x}$, 
which implies the existence of some $o \in \dom{\I}$ with $\tup{\Int{\I}{x},o} \in \Int{\I}{R}$ and $o \in \Int{\I}{C}$.
$\F'$ was obtained by adding to $\F$ a new node which we denote $y$. 
This node will make explicit in $\F$ the existence of $o$, and we will have that $\Int{\I}{y} = o$, 
so $\I \models \F'$.

The case of the  $\geq$-rule is analogous to the $\exists$-rule, 
since in models of $\F'$ we have that $\Int{\I}{y_i} = o_i$ for $1 \leq i \leq n$, 
where $\set{y_1 \cld y_n}$ are the variables added to $\F$ 
and $o_1 \cld o_n$ denote the elements in $\dom{\I}$ s.t. $\tup{\Int{\I}{x},o_i} \in \Int{\I}{R}$ and $o_i \in \Int{\I}{C}$ for the 
variable $x$ in $\F$ to which the rule was applied.
\end{proof}

Finally, we can prove that the union of models of the forests in $\mathsf{ccf}(\nforests{F}{K}{n})$ is exactly the set of all models of $K$.

\begin{proposition}\label{KmodelssomeF}
For every $\I$ such that  $\I \models K$, 
there is some $\F \in \mathsf{ccf}(\nforests{F}{K}{n})$ with $n \geq 0$
such that $\I \models \F$.
\end{proposition}

\begin{proof}

From Lemmas~\ref{FKiffK} and \ref{FmodelssomeFp}, we have that 
for every $\I$ such that  $\I \models K$, 
there is some $\F \in \nforests{F}{K}{n}$ with $n \geq 0$
such that $\I$ is a model of $\F$.
Now we want to prove that there is some $\F_c \in \mathsf{ccf}(\nforests{F}{K}{n})$
such that $\I \models \F_c$.
Suppose there is an interpretation $\I$ such that $\I$ is a model of some completion forest $\F$ that is not complete. 
Then either it is possible to obtain a new forest $\F'$ such that $\I \models \F'$, 
or none  of the propagation rules can be applied. The latest would imply that either $\F$ was complete, which is a contradiction, or that $\F$ had a clash, which is also a contradiction since $\F$ has a model. 
Hence, while applying the propagation rules, the model will be preserved until some complete forest $\F_c$
is reached.  
\end{proof}

%% file: tableExpRules.tex
\small{
\begin{tabular}{|lll|}
\hline
	$\AND$-rule: 		&	if 				& $C_1 \AND C_2 \in \labelfunc{x}$, $x$ is not indirectly blocked \\
									&						& and $\set{C_1, C_2} \nsubseteq \labelfunc{x}$\\
									& then 			& $\tab{L}(x) := \labelfunc{x} \cup \set{C_1, C_2}$\\
\hline									
	$\OR$-rule: 		&	if 				& $C_1 \OR C_2 \in \labelfunc{x}$, $x$ is not indirectly blocked \\
									&						& and $\set{C_1, C_2} \cap \labelfunc{x} = \emptyset$\\
									& then 			& $\labelfunc{x} := \labelfunc{x} \cup \set{E}$ for some $E \in \set{C_1, C_2}$\\
\hline									
	$\exists$-rule: 	&	if 			& $\SOME{S}{C} \in \labelfunc{x}$, $x$ is not blocked and \\
									&						& $x$ has no $S$-neighbour $y$ with $C \in \labelfunc{y}$\\
									& then 			& create new node $y$ with $\labelfunc{\tup{x,y}} := \set{S}$ \\
									&						& and $\labelfunc{x} := \set{C} \cup \axioms{K, \C_K}$\\
\hline									
	$\forall$-rule: 	&	if 			& $\ALL{S}{C} \in \labelfunc{x}$, $x$ is not indirectly blocked and \\
									&						& there is an $S$-neighbour $y$ of $x$ with $C \notin \labelfunc{y}$\\
									& then 			& $\labelfunc{y} := \labelfunc{y} \cup \set{C}$\\
\hline									
	$\forall_{+}$-rule: &	if 		& $\ALL{S}{C} \in \labelfunc{x}$, $x$ is not indirectly blocked, \\
									&						& there is some $R$ with $\ssfunc{Trans}{R}$ and $R \SUBSTRANS S$ and\\
									&						& there is an $S$-neighbour $y$ of $x$ with $\ALL{R}{C} \notin \labelfunc{y}$\\
									& then 			& $\labelfunc{y} := \labelfunc{y} \cup \set{\ALL{R}{C}}$\\
\hline									
	\emph{choose}-rule: 		&	if 				& $\ATMOST{n}{S}{C} \in \labelfunc{x}$ or $\ATLEAST{n}{S}{C} \in \labelfunc{x}$, \\
									&						& $x$ is not indirectly blocked and \\
									&						& there is an $S$-neighbour $y$ of $x$ with $\set{C, NNF(\NOT C)} \cap \labelfunc{y} = \emptyset$\\
									& then 			& $\labelfunc{y} := \labelfunc{y} \cup \set{E}$ for some $E \in \set{C, NNF(\NOT C)}$\\						
\hline									
	$\geq$-rule: 		&	if 				& $\ATLEAST{n}{S}{C} \in \labelfunc{x}$, $x$ is not blocked and\\
									&						& there are not $S$-neighbours $y_1 \cld y_n$ of $x $ such that $C \notin \labelfunc{y_i}$\\
									&						&  and $y_i \NEQ y_j$ for $1 \leq	 i < j \leq n$\\
									& then 			& create new nodes $y_1 \cld y_n$  with $\labelfunc{\tup{x,y_i}} := \set{S}$, \\
									&						& $\labelfunc{y_i} := \set{C} \cup \axioms{K, \C_K}$ and $y_i \NEQ y_j$ for $1 \leq	 i < j \leq n$\\
\hline									
		$\leq$-rule:  &	if 				& $\ATMOST{n}{S}{C} \in \labelfunc{x}$, \\
									&						& $x$ is not indirectly blocked, \\
									&						& $\card{\set{y \st y \text{ is an $S$-neighbour of } x \text{ and } C \in \labelfunc{y}}} > n$ and\\
									&						& there are $S$-neighbours $y$, $z$ of $x$ with not $y \NEQ z$, \\
									&						& $y$ is neither a root node nor an ancestor of $z$ \\
									&						& and $C \in \labelfunc{y} \cap \labelfunc{z}$\\		
									& then 			& $\labelfunc{z} := \labelfunc{z} \cup \labelfunc{y}$, \\
									&						& if $z$ is an ancestor of $x$, \\
									&						& \ \ \ then $\labelfunc{\tup{z,x}} := \labelfunc{\tup{z,x}} \cup \inverse{\labelfunc{\tup{x,y}}}$, \\
									&						& \ \ \ else $\labelfunc{\tup{x,z}} := \labelfunc{\tup{x,z}} \cup \labelfunc{\tup{x,y}}$, \\
									&						& $\labelfunc{\tup{x,y}} := \emptyset$, \\
									&						& set $u \NEQ z$ for all $u$ with  $u \NEQ y$\\							
\hline											
		$\leq_r$-rule:  &	if 				& $\ATMOST{n}{S}{C} \in \labelfunc{x}$, \\
									&						& $\card{\set{y \st y \text{ is an $S$-neighbour of } x \text{ and } C \in \labelfunc{y}}} > n$ and\\
									&						& there are $S$-neighbours $y$, $z$ of $x$ with not $y \NEQ z$, \\
									&						& both $y$ and $z$ are root nodes and $C \in \labelfunc{y} \cap \labelfunc{z}$\\		
									& then 			& $\labelfunc{z} := \labelfunc{z} \cup \labelfunc{y}$, $\labelfunc{y} := \emptyset$ \\
									&						& for all edges $\tup{y,w}$ \\
									&						& $\labelfunc{\tup{z,w}} := \labelfunc{\tup{z,w}} \cup \labelfunc{\tup{y,w}}$, $\labelfunc{\tup{y,w}} := \emptyset$ \\ 
									&						& for all edges $\tup{w,y}$ \\
									&						& 6$\labelfunc{\tup{w,z}} := \labelfunc{\tup{w,z}} \cup \labelfunc{\tup{w,y}}$, $\labelfunc{\tup{w,y}} := \emptyset$ \\
									&						& set $y \EQ z$ and $u \NEQ z$ for all $u$ with  $u \NEQ y$	\\
\hline				
\end{tabular}}

%% file: tableau.tex
\subsection{Tableaux and Canonical Models}

We will define a tableau for a knowledge base. 
A tableau is only a representation of a model of a knowledge base, however, if may be infinite. 
Intuitively, a tableau is a model captured by a complete and clash free completion forest $\F$ and it will provide a natural way of building a canonical interpretation of $\F$.
Note that if $\F$ contains blocked nodes, then it is capturing a set of potentially infinite models. In this case, its tableau must be an infinite structure. 
The tableau $T$ of a forest $\F$ will correspond to the \emph{unraveling} of $\F$. i.e. the structure obtained by considering each path to a node in $\F$ as a node of $T$. 
Following~\cite{indSHIQ}, we will give a rather complex definition of a tableau. 
Defining a model of $K$ from a tableau will be straightforward with this definition, and the many 
conditions required for a tableau are met by complete and crash free completion forests. 

\begin{definition}[Tableau]
$T=\tup{\tabs{S},\tab{L},\tab{E},\tab{I}}$ is a tableau for 
a knowledge base  $K = \tup{\krbox{A}, \krbox{R}, \krbox{T}}$ iff
\begin{itemize}
\item
$\tabs{S}$ is a non-empty set,
\item
$\tab{L} : \tabs{S} \ra 2^{\clos{K}}$ 
maps each element in $\tabs{S}$ to a set of concepts,
\item
$\tab{E}: \roles{K} \ra 2^{\tabs{S} \times \tabs{S}}$ 
maps each role to a set of pairs of elements in $\tabs{S}$, and
\item
$\tab{I}: \indivs{K} \ra  \tabs{S}$ 
maps each individual occurring in $\krbox{A}$ to an element in $\tabs{S}$.
\end{itemize}
Furthermore, for all
$s,t \in \tabs{S}$; 
$C,C_1,C_2 \in \clos{K}$ 
and $R,S \in \roles{K}$,
$T$ satisfies:
    \begin{enumerate}[(P1)]
	\item \label{p1}
	   if $C\in \tab{L}(s)$, then $\NOT C \notin \tab{L}(s)$,
	\item 
	   if $C_1 \AND C_2 \in \tab{L}(s)$, then $C_1 \in \tab{L}(s)$ and $C_2 \in \tab{L}(s)$,
	\item 
	   if $C_1 \OR C_2 \in \tab{L}(s)$, then $C_1 \in \tab{L}(s)$ or $C_2 \in \tab{L}(s)$,
	\item 
	   if $\ALL{S}{C} \in \tab{L}(s)$ and $\tup{s,t} \in \tab{E}(S)$, then $C \in \tab{L}(t)$,
	\item 
	   if $\SOME{S}{C} \in \tab{L}(s)$, then there is some $t \in \tabs{S}$ such that $\tup{s,t} \in \tab{E}(S)$ and $C \in \tab{L}(t)$,
	\item 
	   if $\ALL{S}{C} \in \tab{L}(s)$ and $\tup{s,t} \in \tab{E}(R)$ for some $R \SUBSTRANS S$ with $\ssfunc{Trans}{R} = true$ then $\ALL{S}{C} \in \tab{L}(t)$,
	\item 
		 $\tup{s,t} \in \tab{E}(R)$ iff $\tup{t,s} \in \tab{E}(\mathsf{Inv}(R))$,
	\item 
	   if $\tup{s,t} \in \tab{E}(R)$ and $R \SUBSTRANS S$ then $\tup{s,t} \in \tab{E}(S)$,
	\item 
	   if $\ATMOST{n}{S}{C} \in \tab{L}(s)$, then $\card{\set{t \in \tabs{S} \st \tup{s,t} \in \tab{E}(S) \text{ and } C \in \tab{L}(t)}} \leq n$,
	\item 
	   if $\ATLEAST{n}{S}{C} \in \tab{L}(s)$, then $\card{\set{t \in \tabs{S} \st \tup{s,t} \in \tab{E}(S) \text{ and } C \in \tab{L}(t)}} \geq n$,
	\item 
	   if $\tup{s,t} \in \tab{E}(R)$ and either 
	   $\ATMOST{n}{S}{C} \in \tab{L}(s)$ or $\ATLEAST{n}{S}{C} \in \tab{L}(s)$, 
	   then $C \in \tab{L}(t)$ or $NNF(\NOT C) \in \tab{L}(t)$, 
	\item 
	   if $C(a) \in \krbox{A}$ then $C \in \tab{L}(\tab{I}(a))$,
	\item 
	   if $R(a,b) \in \krbox{A}$ then $\tup{\tab{I}(a), \tab{I}(a)} \in \tab{E}(R)$,
	\item 
	   if $a \neq b \in \krbox{A}$ then $\tab{I}(a) \neq \tab{I}(a)$,
	\item \label{p15}
	   if $C \in \axioms{K,\C}$, then for all $s \in \tabs{S}$ $C\in \tab{L}(s)$. 
    \end{enumerate}
\end{definition}

Trivially, we can obtain a canonical model of a knowledge base from a tableau for it.

\begin{definition}[Canonical Model of a Tableau]
Let $T$ be a tableau.
The \emph{canonical model of $T$}, ${\I}_T = \inter{{\I}_T}$ is defined as follows:
\begin{center}
	$\dom{\I_T} := \tabs{S}$ \\ 
\end{center}
for all concept names $A$ in $\clos{K}$,
\begin{center}
	$\Int{\I_T}{A} := \set{s \st A \in \tab{L}(s)}$
\end{center}
for all individual names $a$ in $\indivs{K}$,
\begin{center}
	$\Int{\I}{a} :=  a$\\
\end{center}
for all role names $R$ in $\krbox{R}$, 

\begin{center}
	$ R^{\I_T}:= \valclos{\tab{E}(R)}$\\
\end{center}
\end{definition}
where $\valclos{\tab{E}(R)}$ the \emph{closure of the extension} of $R$ under $\krbox{R}$, which is defined as:
\begin{center}
 	$ \valclos{\tab{E}(R)}:=	\left\{\
			    \begin{array}{ll}
				(\tab{E}(R))^{+}  		& \text{if $\mathsf{Trans}(R)$}\\
				\tab{E}(R) \cup \mathsf{sub}(\valclos{\tab{E}(R)})	& otherwise
			    \end{array}			
			\right .$	
\end{center}
where $(\tab{E}(R))^{+}$ denotes the transitive closure of $\tab{E}(R)$ 
and 
\begin{displaymath}
	\mathsf{sub}(\valclos{\tab{E}(R)}) = \bigcup_{P \SUBSTRANS R,P\neq R}\valclos{\tab{E}(P)}.
\end{displaymath}

\begin{lemma}
Let $T$ be a tableau for $K$.
The canonical model of $T$ is a model of $K$.
\end{lemma}
\begin{proof}
That $\I_{T}$ is a model of $\krbox{R}$ and $\krbox{A}$ can be proved exactly as in the proof of Lemma~2 in \cite{indSHIQ}. Due to $(P15)$, it can be easily verified that $\I_{T}$ is also a model of $\krbox{T}$.
\end{proof}

\subsubsection{Canonical Interpretation of a Completion Forest.}

A completion forest $\F$ induces a tableau $T_{\F}$, and this tableau gives us a canonical model for $\F$. 

\begin{definition}[Tableau induced by a completion forest]
A \emph{path} in a completion forest $\F$ is a sequence of nodes of the form $p = \path{\pn{x_0}{x'_0} \cld \pn{x_n}{x'_n}}$. In such a path, we define $\tail{p} = x_n$ and $\tailp{p} = x'_n$;
and $\path{p \st \pn{x_{n+1}}{x'_{n+1}}}$ denotes the path $\path{\pn{x_0}{x'_0} \cld \pn{x_n}{x'_n},\pn{x_{n+1}}{x'_{n+1}}}$.
For any path $p$ and variable $z$, 
if $z$ is not blocked and $z$ is an $R$-successor of $\tail{p}$, then $\path{p \st \pn{z}{z}}$ is an \emph{$R$-step} of $p$.
If $z'$ is blocked by $z$ and $z'$ is an $R$-successor of $\tail{p}$, then $\path{p \st \pn{z}{z'}}$ is an $R$-step of $p$. 
If $q$ is an $R$-step of $p$ for some role $R$, then $q$ is a \emph{step} of $p$ and $p$ is a \emph{prefix} of $q$.
The transitive closure of prefix is called subpath.

Given a completion forest $\F$, the set $\paths{F}$ is defined inductively as follows:
\begin{itemize}
    \item
	If $x_0^i$ is a root in $\F$, $\path{\pn{x_0^i}{x_0^i}} \in \paths{F}$.
    \item
	If $p \in \paths{F}$ and $q$ is a step of $p$, then $q \in \paths{F}$.
\end{itemize}
	
The tableau $T_\F = (\tabs{S}, \tab{L}, \tab{E}, \tab{I})$ induced by the completion forest $\F$ is defined as follows:\\
\begin{displaymath}
	\begin{array}{rcl}
		\tabs{S} 	& = 	& \paths{F} \setminus \set{p \st p \in \paths{F} \text{ and } p = \path{\pn{x}{x}} \text{ for some } x \text{ with } \labelfunc{x} = \emptyset}\\
		\tab{L}(p) 	& = 	& \labelfunc{\tail{p}}\\ 

		\tab{E}(R)  	& = 	& \set{\tup{p,q} \in \tabs{S} \times  \tabs{S} \st \text{$q$ is an $R$-step of } p} 	\cup \\
				& 	& \set{\tup{p,q} \in \tabs{S} \times  \tabs{S} \st \text{$p$ is an $\inverse{R}$-step of } q} 	\cup \\
				& 	& \set{\tup{\path{\pn{x}{x}},\path{\pn{y}{y}}} \in \tabs{S} \times  \tabs{S} \st \text{$x$, $y$ are root nodes and $x$ is an $R$-neighbour of $y$}} 
	\end{array}
\end{displaymath}
\end{definition}

\begin{lemma}
Every $\F \in \mathsf{ccf}(\nforests{F}{K}{n})$ for $n \geq 1$ induces a canonical model $\I_{\F}$ for $K$.
\end{lemma}

\begin{proof}
First, it is proved as in \cite{indSHIQ} that every $\F \in \mathsf{ccf}(\nforests{F}{K}{n})$ for $n \geq 1$ induces 
a tableau  $T_{\F}$ for $K$. For the last item of the proof of $(P9)$, note that since $n \geq 1$, pairwise blocking is subsumed and the existence the $u$ predecessor can be ensured.
$(P15)$ also holds due to the following facts:
\begin{itemize}
\item
All nodes $x$ are initialized with $\axioms{K,\C_K} \subseteq \tab{L}(x)$.
\item
The concept names in $\axioms{K,\C_K}$ are never removed from the label of a node unless the label is set to $\emptyset$ by the $\leq_r$-rule. In this case, the label of the node is never modified again.
\end{itemize}
Since $T_{\F}$ is a tableau  for $K$, it has a canonical model $\I_{\F}$ that is a model of $K$. 
The canonical model of $\F$ is $\I_{\F}$ . 
\end{proof}

%% file: conjQuer.tex
\section{Answering Conjunctive Queries}

For a knowledge base $K$ and a query $Q$, we say that $K \models Q$ iff for every interpretation $\I$, $\I \models K$ implies $\I \models Q$.
Analogously, for a completion forest $\F$ and a query $Q$, we say that $\F \models Q$ iff for every interpretation $\I$, $\I \models \F$ implies $\I \models Q$.
We are interested in solving the conjunctive query entailment problem. However, a knowledge base $K$ has an infinite number of possibly infinite models.
The problem is then how to verify that the query $Q$ is entailed by all of them. 
The key issue is that for a given $Q$ it is sufficient to consider the set of complete and clash free $N$-completion forests for $K$, where $N$ is a number that depends on $Q$.
Then, we only have to verify a finite number of structures, all of them of finite size. 
In order to provide a sound and complete algorithm for answering conjunctive queries, we have to prove the following:
\begin{enumerate}[I.]
	\item \label{algCorr1}
		If $K \models Q$ then for every $\F \in \mathsf{ccf}(\nforests{F}{K}{N})$ we can find a mapping from the variables in $Q$ to the variables in $\F$ that witnesses the entailment of the query.
	\item \label{algCorr2}
		If $K$ does not entail $Q$, then there will be some $\F \in \mathsf{ccf}(\nforests{F}{K}{N})$ into which $Q$ can not be mapped. 
\end{enumerate}

From~\ref{algCorr1} and~\ref{algCorr2}, we have an algorithm for checking conjunctive query entailment that works as follows:
an initial completion forest for $K$ is built and expanded using a suitable $N$-blocking as termination condition. Then $Q$ is entailed by $K$ iff the query can be mapped into every complete and clash free completion forest obtained. 

In the following , we will use $Q$ to denote a conjunctive query.
We say that $Q$ can be mapped into a completion forest $\F$, denoted $\models_{\F} Q$, 
if there is a mapping $\sigma:V_Q \ra V_{\F}$ that is the identity mapping for all constants in $V_Q$ and that satisfies the following:
    \begin{enumerate}
	\item\label{cond1map}
	For all $C(x) \in L_Q$, $C \in \labelfunc{\sigma(x)}$.
	\item\label{cond2map}
	For all $R(x,y) \in L_Q$, 
	$\sigma(y)$ is an $R$-descendant of $\sigma(x)$.
    \end{enumerate}

We have already proved that every model of $K$ is a model of some $\F \in \mathsf{ccf}(\nforests{F}{K}{N})$. Hence, if $K \nvDash Q$, then $\F \nvDash Q$ for some $\F$. To prove~\ref{algCorr2}, we only need to prove that if this is the case, then there is no mapping $\sigma$. This is done in the next lemma, which stated that the existence of $\sigma$ suffices to ensure that $\I \models Q$ for every $\I$ model of $\F$. 

\begin{lemma} \label{lemmaifforestthenallmodels}
If $\models_{\F} Q$, then $\F \models Q$.
\end{lemma}

\begin{proof}
Since $\models_{\F} Q$, 
there is a mapping $\sigma:V_Q \ra V_{\F}$ satisfying conditions~\ref{cond1map} and ~\ref{cond2map}.
Take any arbitrary model $\I = \inter{I}$ of $\F$. By definition, it satisfies the following:
    \begin{itemize}
	\item
	if $C \in \labelfunc{x}$, then $\Int{I}{x} \in \Int{I}{C}$
	\item
	if $x$  is an $R$-descendant of $y$, then $\tup{\Int{I}{x}, \Int{I}{y}} \in \Int{I}{R}$.
	\item
	if $x \NEQ y \in \F$, then  $\Int{I}{x} \neq  \Int{I}{y}$
    \end{itemize}  

We can define a mapping $\phi$ from the variables in $V_Q$  to objects in $\dom{I}$ as $\phi(x) = \Int{I}{\sigma(x)}$, and this mapping satisfies $\phi(\vect{Y}) \in \Int{I}{p}$ for all $p(\vect{Y}) \in L_Q$.
\end{proof}

The next step is to prove~\ref{algCorr1}. 
We know that if $K \models Q$, then $\I \models Q$ for any model $\I$ of any $\F \in \mathsf{ccf}(\nforests{F}{K}{N})$. 
We only need to ensure that if this is the case, then the mapping $\sigma$ can be found in $\F$, i.e. we want to consider a suitable $N$ such that the set of complete and clash-free $N$-completion forests can witness on their own the entailment of the query. It suffices to prove that if there is model of $\F$ that is a model of $Q$, then $Q$ can be mapped into $\F$. In particular, we will see that if the canonical model of a forest entails $Q$, then a mapping of $Q$ into $\F$ exists. 

In this proof, the value of $N$ (and hence the termination condition) will play a crucial role. As we mentioned, it depends on $Q$. 
More specifically, it depends in what we call \emph{maximal $Q$-distance}. 
If the canonical model of a forest $\F$ entails $Q$, then there is a mapping of the variables in $Q$ onto the nodes of the tableau induced by $\F$. Intuitively, the \emph{maximal $Q$-distance} is the length of the longest path between two connected nodes of the graph defined by the image of the query when mapped on the tableau.
For a maximal $Q$-distance of $d$ 
it will be possible to find a mapping in an $d$-completion forest that is isomorphic to the image of the query under $\sigma$, since this image does not contain any path of length greater than $d$.
For this reason, we will use the maximal $Q$-distance as blocking condition when expanding the completion forest.

\begin{sloppypar}
Formally, for a given forest $\F$ in $\mathsf{ccf}(\nforests{F}{K}{n})$ for some $n$, let $T_{\F} =\tup{\tabs{S},\tab{L},\tab{E},\tab{I}}$ denote its tableau and $\I_{\F}$ the canonical interpretation of $T_{\F}$. 
If $\I_{\F} \models Q$, then there is a mapping $\sigma: V_Q \ra \tabs{S}$ such that for every $R(x,y) \in L_Q$, $\tup{\sigma(x), \sigma(y)} \in \valclos{\tab{E}(R)}$. 
For each such $R(x,y) \in L_Q$, we use $d^{R}(\sigma(x),\sigma(y))$ to denote the length of the shortest path from $\sigma(x)$ to $\sigma(y)$ in the graph $\tup{\tabs{S}, \bigcup_{P \SUBSTRANS R}\tab{E}(P)}$ and call it the $R$-distance between $\sigma(x)$ and $\sigma(y)$.
For any $x$, $y$ in $V_Q$, $d^{Q}(x,y)$ is the maximal $d^{R}(\sigma(x),\sigma(y))$ that is defined for all $R$ (and it is $0$ if it is not defined for any $R$). 
Let $p$ be a path in the graph $G(Q) = \tup{V_Q,\set{\tup{x,y}\st R(x,y) \in L_Q, R \in \roles{K}}}$, then $d^Q(p) = \sum_{\tup{x,y} \in p} d^{Q}(x,y)$, and 
\begin{displaymath}
	maxd^{Q}(x,y) = max\set{d^{Q}(p) \st p \text{ is a path from $x$ to $y$ in $G(Q)$}}
\end{displaymath}
Finally, the \emph{maximal $Q$-distance}, denoted $d_Q$, is the maximal $maxd^{Q}(x,y)$ that is defined for all $x$, $y$ in $V_Q$, and it is zero if it is not defined for all $x$, $y$. The maximal $Q$-distance is bounded by the length of the longest path in $G(Q)$ (which is bounded by $n_Q$) times the maximal $d^{Q}(x,y)$ that is defined for all $x$, $y$ in $V_Q$.

Now we prove that for any complete and crash free $d_Q$-completion forest $\F$, if $\I_{\F} \models Q$, then there is a mapping $\sigma': V_Q \ra \F$ that witnesses the entailment of $Q$. 
\end{sloppypar}

\begin{proposition}\label{canonicalmodelsQ}
Consider any $\F \in \mathsf{ccf}(\nforests{F}{K}{d_Q})$, and 
let $\I_{\F}$ be the canonical model of the tableau induced by $\F$.
If $\I_{\F} \models Q$ then $\models_{\F} Q$.
\end{proposition}

\begin{proof}
Since $\I_{\F} \models Q$, then there is a mapping $\sigma:V_Q \ra \dom{\I_{\F}}$  s.t.
    \begin{itemize}
	\item
	For all $C(x) \in L_Q$, $\sigma(x) \in \Int{\I_{\F}}{C}$.
	\item 
	For all $R(x,y) \in L_Q$, $\tup{\sigma(x),\sigma(y)} \in \Int{\I_{\F}}{R}$.
    \end{itemize}
Since $\dom{\I_{\F}} = V_{T_{\F}}$, $\sigma(x)$ and $\sigma(y)$ are nodes in $T_{\F}$ and correspond to paths in $\F$.
By the definition of $\I_{\F}$, the mapping $\sigma$ satisfies that for all $C(x) \in L_Q$, $C \in \tab{L}(\sigma(x))$ and for all $R(x,y) \in L_Q$, $\tup{\sigma(x),\sigma(y)} \in \valclos{\tab{E}(R)}$.

We will define a new mapping $\sigma':V_Q \ra V_{\F}$. 
In order to define  $\sigma'$, we will first consider the pairs of variables that are mapped by $\sigma$ to nodes in the  forest such that the path connecting them goes through a leaf of a blocked tree. The set of this pairs will be denoted $\ssfunc{throughLeaves}{V_Q}$.
	For each $R(x,y) \in L_Q$,  if there is some $s \in \tabs{S}$ s.t. $\tup{\sigma(x), s} \in \valclos{\tab{E}(R)}$, $\tup{s, \sigma(y)} \in \valclos{\tab{E}(R)}$ and $\tail{s} \neq \tailp{s}$, then $\tup{x,y} \in \ssfunc{throughLeaves}{V_Q}$.
	The set $\ssfunc{afterblocked}{V_Q}$ will contain the variables in $V_Q$ that occur in the second position of some pair in $\ssfunc{throughLeaves}{V_Q}$ or that are mapped to a descendant of one such node. 
	If $\tup{x,y} \in \ssfunc{throughLeaves}{V_Q}$ or if 
	$R(x,y) \in L_Q$ and $x \in \ssfunc{afterblocked}{V_Q}$, then $y \in \ssfunc{afterblocked}{V_Q}$.
For all variables $v$ in $V_Q \setminus \ssfunc{afterblocked}{V_Q}$, if $\tailp{\sigma(v)}$ is tree blocked let $\psi(\tailp{\sigma(v)}) = \tail{\sigma(v)}$ denote the variable that tree blocks it. Otherwise, let $\psi$ be the identity function.
The mapping $\sigma':V_Q \ra V_{\F}$ is defined as follows:
    \begin{displaymath}
	\sigma'(x)  = \left\{\
	\begin{array}{ll}
		\tailp{\sigma(x)}       & \text{  if } x \in \ssfunc{afterblocked}{V_Q}\\
		\psi(\tailp{\sigma(x)}) & \text{  otherwise}
	\end{array}
	\right .
    \end{displaymath}
Now we will show that the mapping $\sigma'$ has the following properties:
\begin{enumerate}
\item\label{conc}
If $C \in \tab{L}(\sigma(x))$, then $C(x) \in L_Q$, $C \in \L((\sigma'(x)))$.
\item\label{role} 
If  $\tup{\sigma(x),\sigma(y)} \in \valclos{\tab{E}(R)}$, then $\sigma'(y)$ is an $R$-descendant of $\sigma'(x)$.
\end{enumerate}

The proof of~\ref{conc} is trivial, since $\tab{L}(\sigma(x)) =  \labelfunc{\tailp{\sigma(x)}} =  \labelfunc{\psi(\tailp{\sigma(x)})}$, 
so $\tab{L}(\sigma(x)) = \labelfunc{\sigma'(x)}$.
To prove~\ref{role}, first we see that the following hold: 
\begin{description}
	\item[(*)] \label{noLeafsAfter}
	If both $x$ and $y$ are in $\ssfunc{afterblocked}{V_Q}$ and $\tup{\sigma(x),\sigma(y)} \in \valclos{\tab{E}(R)}$ then $\tail{\sigma(y)}$ can not be a blocked leaf. \\
	Since $x$ is in $\ssfunc{afterblocked}{V_Q}$, then by definition there must be some $z \in V_Q$ such that there is a path from $\sigma(z)$ to $\sigma(x)$ in the image of the query that goes through a blocked leaf node, and since there is also a path from $\sigma(x)$ to $\sigma(y)$, if $\tail{\sigma(y)}$ was a blocked leaf then there would be a path from $\sigma(z)$ to $\sigma(y)$  that goes through a blocked leaf and finishes in another blocked leaf. Since we used $d_Q$-blocking, the minimal distance between two blocked leaves is $d_Q + 1$, and then the path from $\sigma(z)$ to $\sigma(y)$ would have a length strictly greater than $d_Q$, which is a contradiction. 
	\item[(**)] \label{noLeafsNoAfter}
	If both $x$ and $y$ are not in $\ssfunc{afterblocked}{V_Q}$ and $\tup{\sigma(x),\sigma(y)} \in \valclos{\tab{E}(R)}$ then $\tail{\sigma(x)}$ can not be a blocked leaf. \\
	If $\tail{\sigma(x)}$ is a blocked leaf and $x$ is not in $\ssfunc{afterblocked}{V_Q}$, then $\tup{x,y}$ is in $\ssfunc{throughLeaves}{V_Q}$ by definition, and then $y$ is in $\ssfunc{afterblocked}{V_Q}$.
\end{description}

By the definition of $\valclos{\tab{E}(R)}$ and of $R$-step, $\tup{\sigma(x),\sigma(y)} \in \valclos{\tab{E}(R)}$ implies that $\tailp{\sigma(y)}$ is an $R$-descendant of $\tail{\sigma(x)}$.
We will now prove that if this is the case, then then $\sigma'(x)$ is an $R$-descendant of $\sigma'(y)$.
Note that since $\sigma(y)$ is an $R$-descendant of $\sigma(x)$, it can not be the case that $x$ is in $\ssfunc{afterblocked}{V_Q}$ and $y$ is not. We have the following cases:
	\begin{enumerate}[(a)]
	    \item
		Both $x$ and $y$ are in $\ssfunc{afterblocked}{V_Q}$.\\
		In this case we have that $\sigma'(x)=\tailp{\sigma(x)}$ and $\sigma'(y)=\tailp{\sigma(y)}$.
		By \textbf{(*)}, $\sigma(y)$ is not a blocked leaf, and then from $\tail{\sigma(y)} = \tailp{\sigma(y)}$ we have that $\tail{\sigma(y)}$ is an $R$-descendant of $\tail{\sigma(x)}$, so $\psi(\tail{\sigma(y)}) = \tailp{\sigma(y)}$ is an $R$-descendant of $\psi(\tail{\sigma(x)})=\tailp{\sigma(x)}$ and $\sigma'(y)$ is an $R$-descendant of $\sigma'(x)$ as desired.
	    \item
		Neither $x$ nor $y$ are in $\ssfunc{afterblocked}{V_Q}$.\\ 		
		By \textbf{(**)}, $\sigma(x)$ is not a blocked leaf, so $\tail{\sigma(x)} = \tailp{\sigma(x)}$ an then $\tailp{\sigma(y)}$ is an $R$-descendant of $\tailp{\sigma(x)}$, so $\psi(\tailp{\sigma(y)})$ is an $R$-descendant of $\psi(\tailp{\sigma(x)})$ as desired.
	    \item
		$x$ is not in  $\ssfunc{afterblocked}{V_Q}$, but $y$ is.\\
		In this case we have that $\sigma(x)$ is a blocked leaf and $\tail{\sigma(x)} = \psi(\tailp{\sigma(x)})$, so
		$\tailp{\sigma(y)} = \sigma'(y)$ is an $R$-descendant of $\psi(\tailp{\sigma(x)}) = \sigma'(x)$.
	\end{enumerate}
Since the mapping $\sigma'$ has properties \ref{conc} and \ref{role}, $\models_{\F} Q$.
\end{proof}

In the absence of transitive roles, $d^{R}(\sigma(x),\sigma(y)) = 1$ for every pair of variables $x$, $y$ that appear in some $R(x,y)$ in $Q$, and then the maximal $Q$-distance is bounded by $n_Q$. 
Due to this fact, it is sufficient to consider $n_Q$-blocking as a termination condition when expanding the completion forest.

\begin{corollary}\label{corCanonicalmodelsQnoTrans}
Let $K$ be a knowledge base with $\transroles = \emptyset$.
Consider any $\F \in \mathsf{ccf}(\nforests{F}{K}{n_Q})$, and 
let $\I_{\F}$ be the canonical model of the tableau induced by $\F$.
If $\I_{\F} \models Q$ then $\models_{\F} Q$.
\end{corollary}

In the presence of transitive roles, if does not suffice to consider $n_Q$-blocking as a termination condition.
Since $d^{R}(\sigma(x),\sigma(y))$ may be arbitrarily big for each $R(x,y)$, then also the maximal $Q$-distance is unbounded and an isomorphic mapping may not exist on a structure of bounded depth. However, as we will now show, if a there is some mapping from the query variables into a tableau for $K$ satisfying $Q$, then there is a mapping that also satisfies $Q$ where the maximal $Q$-distance is bound by a number that depends on $K$.
This will allow us to find an isomorphic mapping of the query variables into a completion forest of fixed size. 
We denote by $\conccard$ the cardinality of $\clos{K} \cup \C_K$ and by $\rolecard$ the cardinality of $\roles{K}$. The bound will be given as $D = 2^{2\conccard+\rolecard}$.
We prove that any mapping where the maximal $d^{R}(\sigma(x),\sigma(y))$ that is defined for some $R$, $x$, $y$ exceeds $D$ can be modified into one that does not. 

\begin{lemma}
Consider a tableau $T=\tup{\tabs{S},\tab{L},\tab{E},\tab{I}}$ for $K$.
If there is a mapping $\sigma':V_Q \ra \tabs{S}$ that satisfies  
    \begin{enumerate}
	\item\label{it1}
	For all $C(x) \in L_Q$, $C \in \tab{L}(\sigma(x))$.
	\item\label{it2}
	For all $R(x,y) \in L_Q$, $\tup{\sigma(x),\sigma(y)} \in \valclos{\tab{E}(R)}$.
    \end{enumerate}
then there is a mapping $\sigma':V_Q \ra \tabs{S}$ that also satisfies~\ref{it1} and~\ref{it2}, and that additionally satisfies that for all $R(x,y) \in L_Q$, $d^{R}(\sigma'(x),\sigma'(y)) \leq D$.
\end{lemma}

\begin{proof}
If $\tup{\sigma(x),\sigma(y)} \in \valclos{\tab{E}(R)}$, then there is a sequence of nodes $n_0, \cld n_m$ s.t. $n_0 = \sigma(x)$, $n_m = \sigma(y)$ and for all $0 \leq i \leq m$, $\tup{n_i,n_{i+1}} \in \tab{E}(S)$ for some $S$ subrole of $R$, and $d^{R}(\sigma(x),\sigma(y)) = m$.
We can prove that if $m > D$, then there is a mapping $\sigma_m$ with $d^{R}(\sigma_m(x),\sigma_m(y)) < m$.
Since there are at most $2^{\conccard}$ node labels and $2^{\rolecard}$ arc labels, there are at most 
$D = 2^{2\conccard+\rolecard}$ possible different labellings for a  pair of nodes and an edge. This implies that if $m > D$, there is some node $m'$ in $n_0, \cld n_m$ that had previously occurred with the same predecessor and the same incoming edge, and hence $n_0, \cld n_m$ contains a cycle. In this case we can consider the path $n_0, \cld m'$ and the new mapping is given as $\sigma_m(x) = \sigma_m(x)$, and $\sigma_m(y) = m'$. 
Inductively, we can prove that there is a mapping $\sigma'$ that satisfies $d^{R}(\sigma'(x),\sigma'(y)) \leq D$ for every  $R(x,y) \in L_Q$.
Since $\sigma'$ preserves all the labels in $\sigma$ and all $R$-descendant relations, $\sigma'$ also satisfies~\ref{it1} and~\ref{it2}.
\end{proof}

Now we know that in the presence of transitive roles, since $d^{R}(\sigma(x),\sigma(y))$ is bounded by $D$, the maximal $Q$-distance is bounded by $D n_Q$, so we can use $D n_Q$-blocking as a termination condition when expanding the completion forest.

\begin{corollary}\label{canonicalmodelsQTrans}
Consider any $\F \in \mathsf{ccf}(\nforests{F}{K}{Dn_Q})$, and 
let $\I_{\F}$ be the canonical model of the tableau induced by $\F$.
If $\I_{\F} \models Q$ then $\models_{\F} Q$.
\end{corollary}

Summing up, to solve the conjunctive query entailment problem, it suffices to check for entailment the set of complete and crash free completion forests for $K$, no matter the $n$ that is used as a termination condition. 

\begin{proposition}\label{legalforstsQ}
$K \models Q$ iff $\F \models Q$ for every $\F \in \mathsf{ccf}(\nforests{F}{K}{n})$ for any $n$.
\end{proposition}

\begin{proof}
The only if direction is trivial. 
Consider any $\F \in \nforests{F}{K}{n}$. 
Since any model $\I$ of $\F$ is a model of $K$ by definition,
then $K \models Q$ implies $\F \models Q$.
The if direction can be done by contraposition. 
If $K \nvDash Q$, then there is some model $\I$ of $K$ such that $\I \nvDash Q$. 
By Proposition~\ref{KmodelssomeF}, $\I \models \F$ for some $\F \in \mathsf{ccf}(\nforests{F}{K}{n})$, 
and we have that $\F  \nvDash Q$ for some $\F \in \mathsf{ccf}(\nforests{F}{K}{n})$.
\end{proof}

However, if we choose a suitable $n$-blocking, checking for entailment in all the models of  a completion forest can be reduced to finding a mapping of the query into the completion forest itself. 

\begin{theorem}\label{kiffallcompf}
$K \models Q$ iff $\models_{\F} Q$ for every $\F \in \mathsf{ccf}(\nforests{F}{K}{d_Q})$.
\end{theorem}

\begin{proof}
First we prove that if $K \models Q$ then $\models_{\F} Q$.
Take any arbitrary $\F \in \mathsf{ccf}(\nforests{F}{K}{d_Q})$. 
Since $K \models Q$,  then  $\F \models Q$ (Proposition~\ref{legalforstsQ}). 
In particular, we have that $\I_{\F} \models Q$, where $\I_{\F}$ is the canonical model of the tableau induced by $\F$.
Thus, by Proposition~\ref{canonicalmodelsQ}, $\models_{\F} Q$.

To prove the other direction, observe that  from $\models_{\F} Q$ and Lemma~\ref{lemmaifforestthenallmodels}, 
we have that $\F \models Q$ for every $\F \in \mathsf{ccf}(\nforests{F}{K}{d_Q})$.
Finally, by  Proposition~\ref{legalforstsQ}, $K \models Q$.
\end{proof}

\begin{corollary}\label{kiffallcompfNoTrans}
If $\transroles = \emptyset$ in $K$, then
$K \models Q$ iff $\models_{\F} Q$ for every $\F \in \mathsf{ccf}(\nforests{F}{K}{n_Q})$.
\end{corollary}

\begin{corollary}\label{kiffallcompfTrans}
$K \models Q$ iff $\models_{\F} Q$ for every $\F \in \mathsf{ccf}(\nforests{F}{K}{D n_Q})$.
\end{corollary}

%% file: complexity.tex
\section{Complexity}

In this section, for a knowledge base $K$, we will use 
$\conccard$ to  denote the cardinality of $\clos{K} \cup \C_K$, 
$\rolecard$ the cardinality of $\roles{K}$ and
$\maxnumrest$ the maximum $m$ occurring in a concept of the form $\ATMOST{m}{R}{C}$ or $\ATLEAST{m}{R}{C}$ in $\clos{K} \cup \C_K$.
$\card{\krbox{A}}$ denotes the number of assertions in $\krbox{A}$.
By $\card{K}$ we will denote the total size of the (string encoding the) knowledge base.
Note that $\conccard$, $\rolecard$ and $\maxnumrest$ are linear on $\card{K \cup \C_K}$ assuming unary coding of numbers in number restrictions and constant on $\card{\krbox{A}}$, while $\card{\indivs{K}}$ is linear on both. 
\begin{lemma}
The maximal number $T_{n}$ of non-isomorphic $n$-trees in a completion forest for $K$ is given by
$T_{n} = \bigoh{(2^{2 \conccard} (\conccard \maxnumrest)^{\rolecard})^{(\conccard \maxnumrest \rolecard)^n}}$.
\end{lemma}

\begin{proof}
Since $\labelfunc{x} \subseteq \clos{K} \cup \C_K$, there are at most $2^{\conccard}$ different node labels in a completion forest. 
Each successor of a node can be the root of a tree of depth $(n-1)$. 
considering a single role $R$, if a node $v$ has $x$ $R$-successors, then there is a maximum number of $(T_{n-1})^x$ trees of depth $(n-1)$ rooted at $v$.
A generating rule can be applied to each node at most $\conccard$ times.
Each time it is applied, it generates at most $\maxnumrest$ $R$-successors for each role $R$. 
This gives a bound of $\conccard \maxnumrest$ $R$-successors for each role.
The number of $R$-successors of a node might range from $0$ to  $\conccard \maxnumrest$, 
and for each number of $R$-successors, we have at most $(T_{n-1})^{(\conccard \maxnumrest)}$ trees of depth $(n-1)$.
So, each node can be the root of at most $(\conccard \maxnumrest) (T_{n-1})^{(\conccard \maxnumrest)}$ trees of depth $(n-1)$ if we consider one single role.
Since at most the same number of trees can be generated for every role in $\roles{K}$, there is a bound of
$((\conccard \maxnumrest) (T_{n-1})^{(\conccard \maxnumrest)})^{\rolecard}$ trees of depth $(n-1)$ rooted at each node. 
The number of different roots of an $n$-tree is bounded by $2^{\conccard}$.
We now give an upper bound on the number of non isomorphic $n$-trees as 
    \begin{displaymath}
 	T_{n} = \bigoh{2^{\conccard} ((\conccard \maxnumrest) (T_{n-1})^{(\conccard \maxnumrest)})^{\rolecard}}
    \end{displaymath}
To simplify the notation, let's consider  $x = 2^{\conccard} (\conccard \maxnumrest)^{\rolecard}$ and
$a = \conccard \maxnumrest \rolecard$.
Then we have
    \begin{displaymath}
 	T_{n} = \bigoh{x (T_{n-1})^{a}} = \bigoh{x^{1+a \ld{+} a^{n-1}} (T_{0})^{a^n}} = \bigoh{(x T_{0})^{a^n}}
    \end{displaymath}
The maximal number of trees of depth $0$ is also bounded by $2^{\conccard}$.
Returning to the original notation we get
    \begin{displaymath}
 	T_{n} = \bigoh{(2^{2 \conccard} (\conccard \maxnumrest)^{\rolecard})^{(\conccard \maxnumrest \rolecard)^n}}
    \end{displaymath}
\end{proof}

\begin{corollary}
The maximal number $T_{n}$ of non-isomorphic $n$-trees in a completion forest for $K$ is:\\
- single exponential in $n$\\
- double exponential in $\card{K}$ if $n$ is constant on $\card{K}$\\
- triple exponential in $\card{K}$ if $n$ is single exponential on $\card{K}$.
\end{corollary}

\begin{lemma}
The number of nodes in a completion forest $\F \in \nforests{F}{K}{n}$ is bounded by 
    \begin{displaymath}
	\bigoh{\card{\indivs{K}}(\conccard \maxnumrest \rolecard)^{n(2^{2 \conccard} (\conccard \maxnumrest)^{\rolecard}) ^{(\conccard \maxnumrest \rolecard)^n}}}
    \end{displaymath}
\end{lemma}

\begin{proof}
The claim follows from  the following properties:
    \begin{enumerate}[i)]
	\item
	The outdegree of $\F$ is bounded by $\conccard \maxnumrest \rolecard$. \\
	Nodes are only added to the forest by applying a generating rule.  
	Only concepts of the form $\SOME{R}{S}$ or $\ATLEAST{n}{R}{C}$ trigger the application of a generating rule, and there are at most $\conccard$ such concepts. 
	Each such rule generates at most $\maxnumrest$ successors for each role, and there are $\rolecard$ roles.
	Note that if a node $v$ is identified with another by the $\forall$-rule or the $\forall_{r}$-rule, then the rule application which led to the generation of $v$ will never be repeated~\cite{indSHIQ}.
	\item
	The depth of $\F$ is bounded by $d = (T_{n}+1)n$.\\
	This is due to the fact that there is a maximum of $T_{n}$ non-isomorphic $n$-trees.
	If there was a path of length greater than $(T_{n}+1)n$ to a node $v$ in $\F$, this would imply that $v$ occurred after a sequence of $T_{n}+1$ non overlapping $n$-trees, and then one of them would have been blocked and $v$ would not have been generated. 
	\item
	The number of variables in a variable tree in $\F$ is bounded by $\bigoh{(\conccard \maxnumrest \rolecard)^{d+1}}$.
	\item
	The number of variables in $\F$ is bounded by  $\bigoh{\card{\indivs{K}}(\conccard \maxnumrest \rolecard)^{d+1}}$.
    \end{enumerate}
\end{proof}

\begin{corollary}
If $n$ is constant on $\card{K}$, then 
the maximum number of nodes in a completion forest $\F \in \nforests{F}{K}{n}$ is 
4-exponential on $(\card{K} + n)$,
3-exponential on $\card{K}$,
double exponential on $n$ and
linear in $\card{\krbox{A}}$.  
\end{corollary}

\begin{corollary}
If $n$ is single exponential on $\card{K}$, then 
the maximum number of nodes in a completion forest $\F \in \nforests{F}{K}{n}$ is 
5-exponential on $(\card{K} + n)$,
4-exponential on $\card{K}$,
double exponential on $n$ and
linear in $\card{\krbox{A}}$.  
\end{corollary}

\begin{proposition}\label{complObtainF}
The expansion of $\F_K$ into some $\F \in \nforests{F}{K}{n}$ terminates in time:\\
- nondeterministic 3-exponential on $\card{K}$ if $n$ is constant on $\card{K}$,\\
- nondeterministic 4-exponential on $(\card{K} + n)$ if $n$ is constant on $\card{K}$,\\
- nondeterministic 4-exponential on $\card{K}$ if $n$ is single exponential on $\card{K}$,\\
- nondeterministic 5-exponential on $(\card{K} + n)$ if $n$ is single exponential on $\card{K}$,\\
- nondeterministic double exponential on $n$,\\
- nondeterministic polynomial (linear) in $\card{\krbox{A}}$.  
\end{proposition}

\begin{proof}
Let $M = \bigoh{\card{\indivs{K}}(\conccard \maxnumrest \rolecard)^{n(2^{2 \conccard} (\conccard \maxnumrest)^{\rolecard}) ^{(\conccard \maxnumrest \rolecard)^n}}}$ denote the maximal number of nodes in $\F$.
We will obtain an upper bound of the number of rules that are applied to expand $\F_K$ into $\F$.

    \begin{enumerate}[i)]
	\item
	For a single node $v$, the $\AND$-rule, the $\OR$-rule and the choose-rule can be applied $\bigoh{\conccard}$ times, since they are applied at most once for each concept in $\labelfunc{v}$.
	\item
	For the $\exists$-rule, $\forall$-rule, $\forall_{+}$-rule, $\geq$-rule and $\leq$-rule, the bound on the number of times it can be applied to $v$ is given by the maximal number of successors of $v$, i.e. $\bigoh{\conccard \maxnumrest \rolecard}$.
	\item
	Rules 1 to 8 can be applied at most $\bigoh{M \conccard \maxnumrest \rolecard}$ times to obtain $\F$.
	\item
	The $\leq_{r}$-rule can be applied at most once to each root node in $\F_K$, hence it is bounded by $\card{\indivs{K}}$.
	\item
	The total rule applications required to expand $\F_K$ into $\F$ is  $\bigoh{\card{\indivs{K}} + (M \conccard \maxnumrest \rolecard)}$
    \end{enumerate}
\end{proof}

\subsection{Complexity of answering Conjunctive Queries}

\begin{lemma}\label{checkingF}
For an $\F \in \ssfunc{ccf}{\nforests{F}{K}{n}}$, checking whether $\models_{\F} Q$ can be done in polynomial time.
\end{lemma}

\begin{proof}[Sketch]
	$\krbox{R}$ and $\F$ can be expressed as a relational database. 
	The complexity of verifying whether $\models_{\F} Q$ is the complexity of answering a conjunctive query over a relational database, which can be done in polynomial time~\cite{abiteboul}.
\end{proof}

\begin{theorem}
Let $K$ be a knowledge base with $\transroles = \emptyset$.
The algorithm answers the conjunctive query entailment problem in $\nconexptime{3}$ \wrt the size of $K$.
\end{theorem}

\begin{proof}
As Theorem~\ref{kiffallcompf} states, $K \nvDash Q$ iff there is some $\F \in \ssfunc{ccf}{\nforests{F}{K}{n}}$ such that $\nvDash_{\F} Q$.
Since $K$ does not contain transitive roles, $n = n_Q$ is constant on $\card{K}$, and 
by Proposition~\ref{complObtainF}, this $\F$ can be obtained in time nondeterministic 3-exponential on $\card{K}$.
From this and Lemma~\ref{checkingF}, we have that non-entailment  is in $\nnexptime{3}$ and the claim follows. 
\end{proof}

\begin{theorem}
Let $K$ be a knowledge base.
The algorithm answers the conjunctive query entailment problem in $\nconexptime{4}$ \wrt the size of $K$.
\end{theorem}

\begin{proof}
As Theorem~\ref{kiffallcompf} states, $K \nvDash Q$ iff there is some $\F \in \ssfunc{ccf}{\nforests{F}{K}{n}}$ such that $\nvDash_{\F} Q$.
Since $n = 2^{2{\conccard \rolecard}} n_Q$ is single exponential on $\card{K}$, by Proposition~\ref{complObtainF} $\F$ can be obtained in time nondeterministic 4-exponential on $\card{K}$.
From this and Lemma~\ref{checkingF}, we have that non-entailment  is in $\nnexptime{4}$ and the claim follows. 
\end{proof}

\subsection{Data Complexity}

\begin{theorem}\label{inconp}
The conjunctive query entailment problem over a knowledge base $K$ in any DL from $\dlf{ALE}$ to $\dlf{SHIQ}$ is in $\conp$ \wrt data complexity.
\end{theorem}

\begin{proof}
Once again, by Theorem~\ref{kiffallcompf} we have that $K \nvDash Q$ iff there is some $\F \in \ssfunc{ccf}{\nforests{F}{K}{n}}$ such that $\nvDash_{\F} Q$.
Proposition~\ref{complObtainF} states that this $\F$ can be obtained in time nondeterministic linear in $\card{\krbox{A}}$, and by Lemma~\ref{checkingF} it can be checked in polynomial time, hence non-entailment is in $\np$ in data complexity, and entailment is in $\conp$.
\end{proof}

\begin{theorem}
The conjunctive query entailment problem over a knowledge base $K$ in any DL from $\dlf{ALE}$ to $\dlf{SHIQ}$ is $\conp$-complete \wrt data complexity.
\end{theorem}

\begin{proof}
The first such hardness result was given in~\cite{Scha94b}, where $\conp$-hardness was proved for $\dlf{ALC}$.
In~\cite{calvanese} the same result is given for logics even less expressive than $\dlf{ALE}$.
Membership for $\dlf{SHIQ}$ is proved in Theorem~\ref{inconp}.
\end{proof}